\documentclass[preprint]{ptephy_v1}

\preprintnumber{KANAZAWA-17-03}

\usepackage{amsmath}
\usepackage{amssymb}

\begin{document}

\title{Higher order tensor renormalization group \\ for relativistic fermion systems}

\author{Ryo Sakai}
\affil{Institute for Theoretical Physics, Kanazawa University, Kanazawa 920-1192, Japan \email{sakai@hep.s.kanazawa-u.ac.jp}}

\author{Shinji Takeda}
\affil{Institute for Theoretical Physics, Kanazawa University, Kanazawa 920-1192, Japan \email{takeda@hep.s.kanazawa-u.ac.jp}}

\author{Yusuke Yoshimura}
\affil{Center for Computational Sciences, University of Tsukuba, Tsukuba 305-8577, Japan \email{yoshimur@ccs.tsukuba.ac.jp}}

\begin{abstract}%
  We apply the higher order tensor renormalization group to two and three dimensional relativistic fermion systems on the lattice.
  In order to perform a coarse-graining of tensor networks including Grassmann variables, we introduce Grassmann higher order tensor renormalization group.
  We test the validity of the new algorithm by comparing its results with those of exact or previous methods.
\end{abstract}


\maketitle

\section{Introduction}
\label{sec:INTRODUCTION}

Monte Carlo simulations of lattice gauge theories have been shown successful as a non-perturbative numerical approach
since the celebrated formulation by Wilson~\cite{Wilson:1974sk} and the first simulation by Creutz~\cite{Creutz:1980zw}.
Such a stochastic approach, however, is generally suffering from the sign problem when the Boltzmann weight is complex.
For example, finite quark density systems, $\theta$ term included systems, or chiral gauge theories
are not easily accessible due to the sign problem.
To avoid the problem, one may rely on other methods using a deterministic algorithm, say the tensor renormalization group (TRG)~\cite{Levin:2006jai}.

The basic procedure of the TRG is as follows.
A target quantity which the TRG can compute is a partition function in the path integral representation. 
For that purpose, first of all, one has to rewrite a partition function in terms of a tensor network.
In general, this step can be done by expanding the Boltzmann weight with some proper expansion scheme depending on the type of physical degrees of freedom (non-compact boson fields, compact boson fields, or fermion fields).
And then after carrying out the integration of the original fields,
it turns out that tensors live at lattice sites and the partition function is expressed as a contraction of them.
The number of terms in the summation is exponentially large as a function of the system size.
To avoid such an expensive computational cost, one may rely on a coarse-graining of the tensor network.
In this step, the singular value decomposition is used to reduce the d.o.f of the system
while preserving important information.
By repeating local blocking transformations, 
one can approximately compute a value of the partition function.
Actually, thanks to the coarse-graining algorithm, the cost now is proportional to the logarithm of system size.

Since the original idea of TRG was introduced by Levin and Nave~\cite{Levin:2006jai},
it has been applied to many two dimensional models associated with elementary particle physics:
the lattice $\phi^{4}$ model~\cite{Shimizu:2012wfa}, the lattice Schwinger model~\cite{Shimizu:2014uva,Shimizu:2014fsa}, and the finite density lattice Thirring model~\cite{Takeda:2014vwa}.
In the studies of the relativistic fermion systems, the Grassmann TRG (GTRG) proposed by Gu \textit{et al}.~\cite{Gu:2010yh,Gu:2013gba} was used.
On the other hand, although the original idea of TRG was limited to two dimensional systems, a new coarse-graining method suited for any higher dimensional system was proposed in Ref.~\cite{2012PhRvB..86d5139X}.
The new method uses the higher order singular value decomposition; thus it is called higher order tensor renormalization group (HOTRG).
The authors in Ref.~\cite{2014ChPhL..31g0503W} analyzed the three dimensional Potts models by using the HOTRG and obtained the critical temperature and exponents precisely.
The HOTRG was also applied in two dimensional systems: the XY model~\cite{Yu:2013sbi}, the O(3) model~\cite{Unmuth-Yockey:2014iga}, and the CP(1) model~\cite{Kawauchi:2016xng}.

It is natural to extend the idea of TRG to higher dimensional systems and more complicated systems.
Our final target is lattice QCD which is a four dimensional relativistic field theory of quarks and gluons.
The tensor network representation of non-Abelian gauge theories was already attempted in Ref.~\cite{Liu:2013nsa},
and the HOTRG for boson systems was already done as mentioned above.
Therefore, the final missing piece to achieve the goal is to treat fermions in higher dimensional systems.
In this paper, we formulate HOTRG for fermions applicable to any dimensional system, and
we call this Grassmann higher order tensor renormalization group (GHOTRG).
As a concrete example, we will provide some details of GHOTRG
in a two dimensional system~\footnote{The extension to four dimensional systems is straightforward although both memory and computational costs are extremely demanding.}.

This paper is organized as follows.
In Sec.~\ref{sec:GHOTRG} we introduce GHOTRG for a two dimensional fermion system with a focus on the Grassmann part in the tensor network.
In Sec.~\ref{sec:RESULTS} we show numerical results and compare them with exact or previous ones.
A summary and an outlook for the future work are given in Sec.~\ref{sec:SUMMARY}.

\section{Grassmann higher order tensor renormalization group}
\label{sec:GHOTRG}

In this section, after briefly reviewing the original HOTRG, we explain GHOTRG in two dimensions in detail.
The three dimensional version will be obtained straightforwardly.
Finally we discuss how to treat the anti-periodic boundary conditions for fermion fields.
Lattice units $a=1$ are assumed in the following.

\subsection{Model and notation in two dimensional systems}

The Lagrangian density of the free Wilson fermion~\footnote{
  Inclusion of interaction terms is straightforward as seen in~\cite{Shimizu:2014uva} and~\cite{Takeda:2014vwa}
  for the Schwinger model and the Thirring model, respectively.
} in two dimensions is given by
\begin{align}
  \label{eq:1}
  &\mathcal{L}_{n}=\bar{\psi}_{n}\left(D\psi\right)_{n},\\
  \label{eq:2}
  &D_{n,n^{\prime}}=\left(m+2\right)\delta_{n,n^{\prime}}-\frac{1}{2}\sum_{\nu,\pm}e^{\mp\mu\delta_{\nu,2}}\left(1\pm \gamma_{\nu}\right)\delta_{n,n^{\prime}\pm\hat{\nu}},
\end{align}
where $n=(n_1,n_2)$ is the lattice coordinate, the fermion fields $\psi$, $\bar\psi$ have two spinor components, and $m$ and $\mu$ denote the mass and the chemical potential.
Following the prescription described in Refs.~\cite{Shimizu:2014uva,Takeda:2014vwa}, one can obtain tensor network representation of the partition function 
\begin{align}
  \label{eq:3}
  Z=\int\mathcal{D}\psi\mathcal{D}\bar{\psi}e^{-\sum_{n}\mathcal{L}_{n}}
  =\sum_{\{x,t\}}\int\prod_{n}\mathcal{T}_{x_{n}t_{n}x_{n-\hat{1}}t_{n-\hat{2}}}.
\end{align}
The tensor $\mathcal{T}$, an elementary building block of the tensor network, is defined as
\begin{align}
  \label{eq:FirstT}
  \mathcal{T}_{x_{n}t_{n}x_{n-\hat{1}}t_{n-\hat{2}}}=
  \begin{aligned}[t]
    &T_{x_{n}t_{n}x_{n-\hat{1}}t_{n-\hat{2}}}\mathrm{d}\bar{\eta}_{n,2}^{x_{n,2}}\mathrm{d}\eta_{n,1}^{x_{n,1}}\mathrm{d}\bar{\xi}_{n,2}^{t_{n,2}}\mathrm{d}\xi_{n,1}^{t_{n,1}}\mathrm{d}\eta_{n,2}^{x_{n-\hat{1},2}}\mathrm{d}\bar{\eta}_{n,1}^{x_{n-\hat{1},1}}\mathrm{d}\xi_{n,2}^{t_{n-\hat{2},2}}\mathrm{d}\bar{\xi}_{n,1}^{t_{n-\hat{2},1}}\\
    &\cdot\left(\bar{\eta}_{n+\hat{1},1}\eta_{n,1}\right)^{x_{n,1}}\left(\bar{\eta}_{n,2}\eta_{n+\hat{1},2}\right)^{x_{n,2}}\left(\bar{\xi}_{n+\hat{2},1}\xi_{n,1}\right)^{t_{n,1}}\left(\bar{\xi}_{n,2}\xi_{n+\hat{2},2}\right)^{t_{n,2}},
  \end{aligned}
\end{align}
where the original fermion fields $\psi$, $\bar\psi$ have been already integrated out while the another set of Grassmann variables $\eta_{n,i}$, $\bar{\eta}_{n,i}$, $\xi_{n,i}$, and $\bar{\xi}_{n,i}$ ($i=1,2$) has been introduced. They are independent each other and satisfy
\begin{align}
  \int \mathrm{d}\eta_{n,i}\eta_{n,i}=\int \mathrm{d}\bar{\eta}_{n,i}\bar\eta_{n,i}=\int \mathrm{d}\xi_{n,i}\xi_{n,i}=\int \mathrm{d}\bar{\xi}_{n,i}\bar\xi_{n,i}=1,&&\text{for }i=1,2.
\end{align}
The graphical expression of the tensor is shown in Fig.~\ref{fig:1-direction} (left) together with the index assignment.
In Eq.~(\ref{eq:FirstT}), $T_{x_nt_nx_{n-\hat{1}}t_{n-\hat{2}}}$ is a normal tensor whose components are normal numbers,
and we call it bosonic tensor.
The elements of the bosonic tensor are explicitly given in Refs.~\cite{Shimizu:2014uva,Takeda:2014vwa} for the Schwinger model and the Thirring model, respectively.
A bunch of Grassmann variables in Eq.~(\ref{eq:FirstT})
\begin{align}
  \label{eq:4}
  &\mathrm{d}\bar{\eta}_{n,2}^{x_{n,2}}\mathrm{d}\eta_{n,1}^{x_{n,1}}\mathrm{d}\bar{\xi}_{n,2}^{t_{n,2}}\mathrm{d}\xi_{n,1}^{t_{n,1}}\mathrm{d}\eta_{n,2}^{x_{n-\hat{1},2}}\mathrm{d}\bar{\eta}_{n,1}^{x_{n-\hat{1},1}}\mathrm{d}\xi_{n,2}^{t_{n-\hat{2},2}}\mathrm{d}\bar{\xi}_{n,1}^{t_{n-\hat{2},1}}\\
  &\cdot\left(\bar{\eta}_{n+\hat{1},1}\eta_{n,1}\right)^{x_{n,1}}\left(\bar{\eta}_{n,2}\eta_{n+\hat{1},2}\right)^{x_{n,2}}\left(\bar{\xi}_{n+\hat{2},1}\xi_{n,1}\right)^{t_{n,1}}\left(\bar{\xi}_{n,2}\xi_{n+\hat{2},2}\right)^{t_{n,2}}\nonumber
\end{align}
is called Grassmann part.
Note that the indices of tensor in Eq.~(\ref{eq:FirstT}) should be read as
\begin{align}
  x_n=&(x_{n,1},x_{n,2}),
        \label{eq:x12}
  \\
  t_n=&(t_{n,1},t_{n,2}),
\end{align}
reflecting the fact that the original Grassmann variable $\psi$ has two components.
Each index runs from 0 to 1; thus the tensor
\begin{align}
  \label{eq:5}
  \mathcal{T}_{x_{n}t_{n}x_{n-\hat{1}}t_{n-\hat{2}}}=\mathcal{T}_{(x_{n,1},x_{n,2})(t_{n,1},t_{n,2})(x_{n-\hat{1},1},x_{n-\hat{1},2})(t_{n-\hat{2},1},t_{n-\hat{2},2})}
\end{align}
has $2^{2\times4}$ elements at the initial stage.
Due to the Grassmann nature, the element of the tensor takes a non-trivial value only when the indices satisfy
\begin{align}
  \label{eq:condition1}
  \left[\sum_{i=1}^{2}\left(x_{n,i}+t_{n,i}+x_{n-\hat{1},i}+t_{n-\hat{2},i}\right)\right]\bmod 2=0,&&\text{for all }n,
\end{align}
otherwise the element is zero.

In the following we discuss the renormalization procedure for the bosonic tensor and the Grassmann part separately.

\subsection{Normal HOTRG procedure for the bosonic tensor}
In this subsection, let us see the renormalization of bosonic tensors, especially focusing on the coarse-graining along the $\hat{1}$-direction.
This is a brief review of the normal HOTRG~\cite{2012PhRvB..86d5139X}.

\begin{figure}[htbp]
  \centering
  \includegraphics{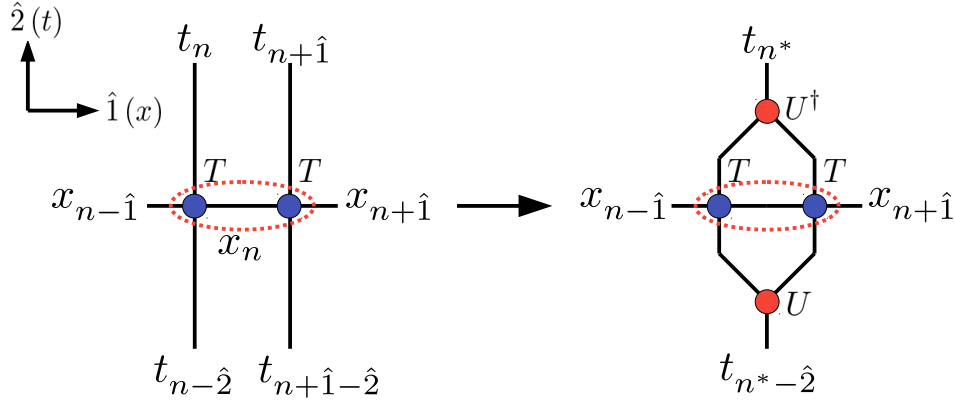}
  \caption{Coarse-graining along the $\hat{1}$-direction.}
  \label{fig:1-direction}
\end{figure}

First we consider the new tensor $M$ by contracting two bosonic tensors placed next to each other along the $\hat{1}$-direction (see Fig.~\ref{fig:1-direction})
\begin{align}
  \label{eq:MakeM}
  &M_{x_{n+\hat{1}}t^{+}_{n}x_{n-\hat{1}}t^{-}_{n}}=\sum_{x_{n}}T_{x_{n}t_{n}x_{n-\hat{1}}t_{n-\hat{2}}}T_{x_{n+\hat{1}}t_{n+\hat{1}}x_{n}t_{n+\hat{1}-\hat{2}}},
\end{align}
where the integrated indices are defined as
\begin{align}
  \label{eq:7}
  &t^{+}_{n}=t_{n}\otimes t_{n+\hat{1}}, \\
  \label{eq:14}
  &t^{-}_{n}=t_{n-\hat{2}}\otimes t_{n+\hat{1}-\hat{2}}.
\end{align}
Then one defines $M^{+}$ and $M^{-}$ as
\begin{align}
  \label{eq:9}
  &M^{\pm}_{t^{\pm}_{n},{t^{\pm}_{n}}^{\prime}}=\sum_{x_{n+\hat{1}},x_{n-\hat{1}},t^{\mp}_{n}}M^{\prime}_{t^{\pm}_{n},x_{n+\hat{1}}x_{n-\hat{1}}t^{\mp}_{n}}M^{\prime\dagger}_{x_{n+\hat{1}}x_{n-\hat{1}}t^{\mp}_{n},{t^{\pm}_{n}}^{\prime}},
\end{align}
with
\begin{align}
  \label{eq:8}
  &M^\prime_{t^{\pm}_{n},x_{n+\hat{1}}x_{n-\hat{1}}t^{\mp}_{n}}= M_{x_{n+\hat{1}}t^{+}_{n}x_{n-\hat{1}}t^{-}_{n}}.
\end{align}
Next we apply the eigenvalue decomposition to $M^{+}$ and obtain a unitary matrix $U^{+}$ and the eigenvalues $\lambda^{+}$: 
\begin{align}
  \label{eq:EVD}
  M^{+}_{t^{+}_{n},{t^{+}_{n}}^{\prime}}=\sum_{t_{n^{*},\mathrm{b}}}U^{+}_{t^{+}_{n},t_{n^{*},\mathrm{b}}}\lambda^{+}_{t_{n^{*},\mathrm{b}}}{U^{+}}^\dagger_{t_{n^{*},\mathrm{b}},{t^{+}_{n}}^{\prime}}, 
\end{align}
where the new index $t_{n^{*},\mathrm{b}}$ will be regarded as the second component of the index of the new tensor $T^{\mathrm{new,b}}$ as it will be clear soon (see Eq.~(\ref{eq:13})).
Similarly one obtains $U^{-}$ and $\lambda^{-}$ from $M^{-}$,
and then we define $\epsilon^{+}$ and $\epsilon^{-}$ as
\begin{align}
  \label{eq:12}
  \epsilon^{\pm}=\sum_{i>D_{\mathrm{cut}}}\lambda^{\pm}_{i},
\end{align}
with a given $D_{\mathrm{cut}}$ one can choose.
$\epsilon^{\pm}$ represent an amount of truncation error and are used to select a unitary matrix which maintains better precision. That is,
if $\epsilon^{+}<\epsilon^{-}$, then $U^{+}$ is adopted and vice versa.
Now one obtains the new tensor by using the selected unitary matrix (denoted by $U$) and restricting the new indices $1\le t_{n^{*},\mathrm{b}}, t_{n^{*}-\hat{2},\mathrm{b}}\le D_{\mathrm{cut}}$,
\begin{align}
  \label{eq:13}
  T^{\mathrm{new,b}}_{x_{n+\hat{1}}t_{n^{*},\mathrm{b}}x_{n-\hat{1}}t_{n^{*}-\hat{2},\mathrm{b}}}=\sum_{t^{+}_{n},t^{-}_{n}}^{\mathrm{all}}U^{*}_{t^{+}_{n}t_{n^{*},\mathrm{b}}}M_{x_{n+\hat{1}}t^{+}_{n}x_{n-\hat{1}}t^{-}_{n}}U_{t^{-}_{n}t_{n^{*}-\hat{2},\mathrm{b}}}.
\end{align}
Actually, when dealing with the Grassmann variables, this part should be modified in order to incorporate the sign factors originated from the coarse-graining of them. See Eq.~(\ref{eq:TnewGrassmann0}) for details.

\begin{figure}[htbp]
  \centering
  \includegraphics{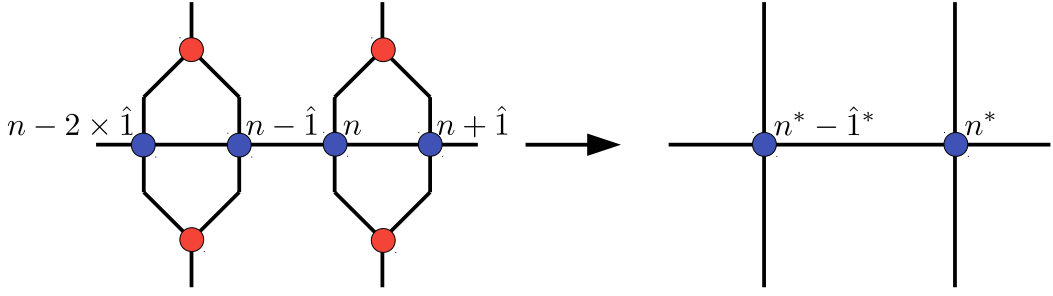}
  \caption{Correspondence between $n$ and $n^{*}$.}
  \label{fig:Rename}
\end{figure}

The indices of the new tensor now contain two kinds of coordinates $n$ and $n^{*}$.
At this point, we make clear the relationship between $n$ and $n^{*}$ as shown in Fig.~\ref{fig:Rename}.
Following the relationship, the indices are changed as follows:
\begin{align}
      T^{\mathrm{new,b}}_{x_{n+\hat{1}}t_{n^{*},\mathrm{b}}x_{n-\hat{1}}t_{n^{*}-\hat{2},\mathrm{b}}}&\rightarrow T^{\mathrm{new,b}}_{x_{n^{*}}t_{n^{*},\mathrm{b}}x_{n^{*}-\hat{1}^{*}}t_{n^{*}-\hat{2},\mathrm{b}}} \nonumber\\
  \label{eq:Rename}
                                                                                                 &\rightarrow T^{\mathrm{new,b}}_{x_{n}t_{n,\mathrm{b}}x_{n-\hat{1}^{*}}t_{n-\hat{2},\mathrm{b}}}.
\end{align}
In the last step, $n^{*}$ has been renamed $n$ since all coordinates are labeled by $n^{*}$.
To remember that the lattice spacing of the $\hat{1}$-direction is doubled, however, the new unit vector defined as $\hat{1}^{*}=\hat{1}+\hat{1}$ will be used.

\subsection{Coarse-graining of the Grassmann part}

In this subsection, we discuss the coarse-graining of the Grassmann part.
Our objective here is to coarse-grain the Grassmann parts of the two tensors
$\mathcal{T}_{x_{n}t_{n}x_{n-\hat{1}}t_{n-\hat{2}}}$
and
$\mathcal{T}_{x_{n+\hat{1}}t_{n+\hat{1}}x_{n}t_{n+\hat{1}-\hat{2}}}$, which are located next to each other along the $\hat{1}$-direction,
and obtain a new Grassmann part with new Grassmann variables.

For that purpose, first of all, we collect their Grassmann parts
\begin{align}
  \label{eq:10}
  &\mathrm{d}\bar{\eta}_{n+\hat{1},2}^{x_{n+\hat{1},2}}\mathrm{d}\eta_{n+\hat{1},1}^{x_{n+\hat{1},1}}\mathrm{d}\bar{\xi}_{n+\hat{1},2}^{t_{n+\hat{1},2}}\mathrm{d}\xi_{n+\hat{1},1}^{t_{n+\hat{1},1}}\mathrm{d}\eta_{n+\hat{1},2}^{x_{n,2}}\mathrm{d}\bar{\eta}_{n+\hat{1},1}^{x_{n,1}}\mathrm{d}\xi_{n+\hat{1},2}^{t_{n+\hat{1}-\hat{2},2}}\mathrm{d}\bar{\xi}_{n+\hat{1},1}^{t_{n+\hat{1}-\hat{2},1}}\\
  &\cdot\mathrm{d}\bar{\eta}_{n,2}^{x_{n,2}}\mathrm{d}\eta_{n,1}^{x_{n,1}}\mathrm{d}\bar{\xi}_{n,2}^{t_{n,2}}\mathrm{d}\xi_{n,1}^{t_{n,1}}\mathrm{d}\eta_{n,2}^{x_{n-\hat{1},2}}\mathrm{d}\bar{\eta}_{n,1}^{x_{n-\hat{1},1}}\mathrm{d}\xi_{n,2}^{t_{n-\hat{2},2}}\mathrm{d}\bar{\xi}_{n,1}^{t_{n-\hat{2},1}}\nonumber\\
  &\cdot\left(\bar{\eta}_{n+\hat{1}+\hat{1},1}\eta_{n+\hat{1},1}\right)^{x_{n+\hat{1},1}}\left(\bar{\eta}_{n+\hat{1},2}\eta_{n+\hat{1}+\hat{1},2}\right)^{x_{n+\hat{1},2}}\left(\bar{\xi}_{n+\hat{1}+\hat{2},1}\xi_{n+\hat{1},1}\right)^{t_{n+\hat{1},1}}\left(\bar{\xi}_{n+\hat{1},2}\xi_{n+\hat{1}+\hat{2},2}\right)^{t_{n+\hat{1},2}}\nonumber\\
  &\cdot\left(\bar{\eta}_{n+\hat{1},1}\eta_{n,1}\right)^{x_{n,1}}\left(\bar{\eta}_{n,2}\eta_{n+\hat{1},2}\right)^{x_{n,2}}\left(\bar{\xi}_{n+\hat{2},1}\xi_{n,1}\right)^{t_{n,1}}\left(\bar{\xi}_{n,2}\xi_{n+\hat{2},2}\right)^{t_{n,2}}.\nonumber
\end{align}
Note that since the indices of tensor satisfy Eq.~(\ref{eq:condition1}),
the Grassmann part of each tensor can freely move around without invoking awkward sign factors. 
Next we can simply integrate out the Grassmann variables $\eta_{n,1}^{x_{n,1}}$, $\bar{\eta}_{n+\hat{1},1}^{x_{n,1}}$, $\eta_{n+\hat{1},2}^{x_{n,2}}$, and $\bar{\eta}_{n,2}^{x_{n,2}}$ since the associated measure is contained in Eq.~(\ref{eq:10}).
This integration corresponds to the summation of $x_{n}$ in Eq.~(\ref{eq:MakeM}) for the bosonic part,
and Eq.~(\ref{eq:10}) turns out to be
\begin{align}
  \label{eq:11}
  &\left(-1\right)^{x_{n,1}\left(x_{n,1}+x_{n,2}+t_{n+\hat{1}-\hat{2},1}+t_{n+\hat{1}-\hat{2},2}\right)+x_{n,2}\left(t_{n+\hat{1}-\hat{2},1}+t_{n+\hat{1}-\hat{2},2}\right)}\\
  &\cdot\mathrm{d}\bar{\eta}_{n+\hat{1},2}^{x_{n+\hat{1},2}}\mathrm{d}\eta_{n+\hat{1},1}^{x_{n+\hat{1},1}}\mathrm{d}\bar{\xi}_{n+\hat{1},2}^{t_{n+\hat{1},2}}\mathrm{d}\xi_{n+\hat{1},1}^{t_{n+\hat{1},1}}\mathrm{d}\xi_{n+\hat{1},2}^{t_{n+\hat{1}-\hat{2},2}}\mathrm{d}\bar{\xi}_{n+\hat{1},1}^{t_{n+\hat{1}-\hat{2},1}}\nonumber\\
  &\cdot\mathrm{d}\bar{\xi}_{n,2}^{t_{n,2}}\mathrm{d}\xi_{n,1}^{t_{n,1}}\mathrm{d}\eta_{n,2}^{x_{n-\hat{1},2}}\mathrm{d}\bar{\eta}_{n,1}^{x_{n-\hat{1},1}}\mathrm{d}\xi_{n,2}^{t_{n-\hat{2},2}}\mathrm{d}\bar{\xi}_{n,1}^{t_{n-\hat{2},1}}\nonumber\\
  &\cdot\left(\bar{\eta}_{n+\hat{1}+\hat{1},1}\eta_{n+\hat{1},1}\right)^{x_{n+\hat{1},1}}\left(\bar{\eta}_{n+\hat{1},2}\eta_{n+\hat{1}+\hat{1},2}\right)^{x_{n+\hat{1},2}}\left(\bar{\xi}_{n+\hat{1}+\hat{2},1}\xi_{n+\hat{1},1}\right)^{t_{n+\hat{1},1}}\left(\bar{\xi}_{n+\hat{1},2}\xi_{n+\hat{1}+\hat{2},2}\right)^{t_{n+\hat{1},2}}\nonumber\\
  &\cdot\left(\bar{\xi}_{n+\hat{2},1}\xi_{n,1}\right)^{t_{n,1}}\left(\bar{\xi}_{n,2}\xi_{n+\hat{2},2}\right)^{t_{n,2}}.\nonumber
\end{align}
The sign factors have arisen along with the interchange of Grassmann variables.

\begin{figure}[htbp]
  \centering
  \includegraphics{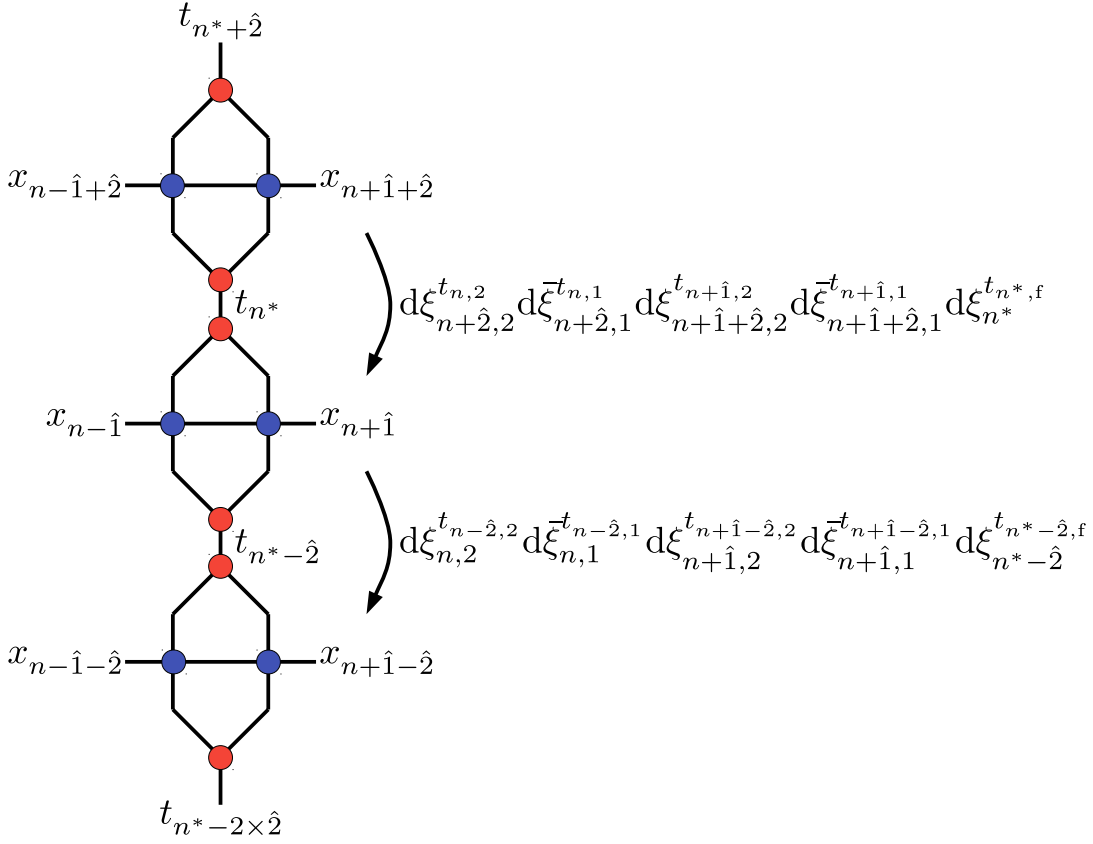}
  \caption{Shift of Grassmann measures.}
  \label{pass}
\end{figure}

The next step is to integrate out the ``old'' Grassmann variables $\xi$ in the hopping factors for the $\hat 1$-direction.
For $\xi_{n,1}^{t_{n,1}}$, $\bar{\xi}_{n,2}^{t_{n,2}}$, $\xi_{n+\hat{1},1}^{t_{n+\hat{1},1}}$, and $\bar{\xi}_{n+\hat{1},2}^{t_{n+\hat{1},2}}$,
one can simply integrate out them since the corresponding measures are already there, and then Eq.~(\ref{eq:11}) turns out to be
\begin{align}
  \label{eq:combi1}
  &\left(-1\right)^{x_{n,1}\left(x_{n,1}+x_{n,2}+t_{n+\hat{1}-\hat{2},1}+t_{n+\hat{1}-\hat{2},2}\right)+x_{n,2}\left(t_{n+\hat{1}-\hat{2},1}+t_{n+\hat{1}-\hat{2},2}\right)+\left(t_{n,1}+t_{n+\hat{1},1}\right)}\\
  &\cdot\mathrm{d}\bar{\eta}_{n+\hat{1},2}^{x_{n+\hat{1},2}}\mathrm{d}\eta_{n+\hat{1},1}^{x_{n+\hat{1},1}}\xi_{n+\hat{1}+\hat{2},2}^{t_{n+\hat{1},2}}\bar{\xi}_{n+\hat{1}+\hat{2},1}^{t_{n+\hat{1},1}}\mathrm{d}\xi_{n+\hat{1},2}^{t_{n+\hat{1}-\hat{2},2}}\mathrm{d}\bar{\xi}_{n+\hat{1},1}^{t_{n+\hat{1}-\hat{2},1}}\nonumber\\
  &\cdot\xi_{n+\hat{2},2}^{t_{n,2}}\bar{\xi}_{n+\hat{2},1}^{t_{n,1}}\mathrm{d}\eta_{n,2}^{x_{n-\hat{1},2}}\mathrm{d}\bar{\eta}_{n,1}^{x_{n-\hat{1},1}}\mathrm{d}\xi_{n,2}^{t_{n-\hat{2},2}}\mathrm{d}\bar{\xi}_{n,1}^{t_{n-\hat{2},1}}\nonumber\\
  &\cdot\left(\bar{\eta}_{n+\hat{1}+\hat{1},1}\eta_{n+\hat{1},1}\right)^{x_{n+\hat{1},1}}\left(\bar{\eta}_{n+\hat{1},2}\eta_{n+\hat{1}+\hat{1},2}\right)^{x_{n+\hat{1},2}},\nonumber
\end{align}
where the additional sign factors arise when breaking the hopping factors for $\xi$.
On the other hand, for
$\bar{\xi}_{n+\hat{2},1}^{t_{n,1}}$, $\xi_{n+\hat{2},2}^{t_{n,2}}$, $\bar{\xi}_{n+\hat{1}+\hat{2},1}^{t_{n+\hat{1},1}}$, and $\xi_{n+\hat{1}+\hat{2},2}^{t_{n+\hat{1},2}}$, 
there are no corresponding measures in Eq.~(\ref{eq:combi1}).
This mismatch, however, can be resolved by noting that the desired measures are obtained by shifting a site to the $\hat2$-direction.
In other words, the corresponding measures~\footnote{Note that the Grassmann part for 
  other sites are there as written in Eq.~(\ref{eq:3}).} exist at neighboring site $n+\hat 2$.
This statement can be easily seen if one arranges the ordering of the Grassmann variables and their measures
\begin{align}
  \label{eq:combi}
  &\left(-1\right)^{x_{n,1}\left(x_{n,1}+x_{n,2}\right)+\left(t_{n,1}+t_{n+\hat{1},1}\right)+t_{n,1}t_{n,2}+t_{n+\hat{1},1}t_{n+\hat{1},2}}\\
  &\cdot\mathrm{d}\bar{\eta}_{n+\hat{1},2}^{x_{n+\hat{1},2}}\mathrm{d}\eta_{n+\hat{1},1}^{x_{n+\hat{1},1}}\bar{\xi}_{n+\hat{1}+\hat{2},1}^{t_{n+\hat{1},1}}\xi_{n+\hat{1}+\hat{2},2}^{t_{n+\hat{1},2}}\bar{\xi}_{n+\hat{2},1}^{t_{n,1}}\xi_{n+\hat{2},2}^{t_{n,2}}\nonumber\\
  &\cdot\mathrm{d}\eta_{n,2}^{x_{n-\hat{1},2}}\mathrm{d}\bar{\eta}_{n,1}^{x_{n-\hat{1},1}}\mathrm{d}\xi_{n,2}^{t_{n-\hat{2},2}}\mathrm{d}\bar{\xi}_{n,1}^{t_{n-\hat{2},1}}\mathrm{d}\xi_{n+\hat{1},2}^{t_{n+\hat{1}-\hat{2},2}}\mathrm{d}\bar{\xi}_{n+\hat{1},1}^{t_{n+\hat{1}-\hat{2},1}}\nonumber\\
  &\cdot\left(\bar{\eta}_{n+\hat{1}+\hat{1},1}\eta_{n+\hat{1},1}\right)^{x_{n+\hat{1},1}}\left(\bar{\eta}_{n+\hat{1},2}\eta_{n+\hat{1}+\hat{1},2}\right)^{x_{n+\hat{1},2}}.\nonumber
\end{align}
Needless to say, when moving a bunch of the measures for $\xi$ in Eq.~(\ref{eq:combi})
\begin{align}
  \label{eq:WantToMove}
  \mathrm{d}\xi_{n,2}^{t_{n-\hat{2},2}}\mathrm{d}\bar{\xi}_{n,1}^{t_{n-\hat{2},1}}\mathrm{d}\xi_{n+\hat{1},2}^{t_{n+\hat{1}-\hat{2},2}}\mathrm{d}\bar{\xi}_{n+\hat{1},1}^{t_{n+\hat{1}-\hat{2},1}},
\end{align}
one should take care of numerous sign factors.
In order to control them, 
we introduce a new index and a new set of Grassmann variables~\footnote{Note that the new Grassmann variable has a single component.} satisfying
\begin{align}
  \label{eq:NewIAndG}
  \left(\mathrm{d}\xi_{n^{*}-\hat{2}}\mathrm{d}\bar{\xi}_{n^{*}}\bar{\xi}_{n^{*}}\xi_{n^{*}-\hat{2}}\right)^{t_{n^{*}-\hat{2},\mathrm{f}}}=1,
\end{align}
where the new index $t_{n^{*}-\hat{2},\mathrm{f}}\in\left\{0,1\right\}$ is defined as
\begin{align}
  \label{eq:condition2}
  t_{n^{*}-\hat{2},\mathrm{f}}=\left(t_{n-\hat{2},1}+t_{n-\hat{2},2}+t_{n+\hat{1}-\hat{2},1}+t_{n+\hat{1}-\hat{2},2}\right)\bmod 2.
\end{align}
Thanks to Eq.~(\ref{eq:condition2}), the combination of Grassmann measures
\begin{align}
  \label{eq:NowGEven}
  \mathrm{d}\xi_{n,2}^{t_{n-\hat{2},2}}\mathrm{d}\bar{\xi}_{n,1}^{t_{n-\hat{2},1}}\mathrm{d}\xi_{n+\hat{1},2}^{t_{n+\hat{1}-\hat{2},2}}\mathrm{d}\bar{\xi}_{n+\hat{1},1}^{t_{n+\hat{1}-\hat{2},1}}\mathrm{d}\xi_{n^{*}-\hat{2}}^{t_{n^{*}-\hat{2},\mathrm{f}}}
\end{align}
is Grassmann-even; thus we can now freely move it. 
By inserting Eq.~(\ref{eq:NewIAndG}) into the proper position in Eq.~(\ref{eq:combi}) and moving the combination of Grassmann measures in Eq.~(\ref{eq:NowGEven}) as shown in Fig.~\ref{pass},
one can execute the integration of the old Grassmann variable $\xi$:
\begin{align}
  \label{eq:BeforeRename}
  &\left(-1\right)^{x_{n,1}\left(x_{n,1}+x_{n,2}\right)+t_{n,2}\left(t_{n,1}+t_{n,2}\right)+t_{n+\hat{1},2}\left(t_{n+\hat{1},1}+t_{n+\hat{1},2}\right)}\\
  &\cdot\mathrm{d}\bar{\eta}_{n+\hat{1},2}^{x_{n+\hat{1},2}}\mathrm{d}\eta_{n+\hat{1},1}^{x_{n+\hat{1},1}}\mathrm{d}\xi_{n^{*}}^{t_{n^{*},\mathrm{f}}}\mathrm{d}\eta_{n,2}^{x_{n-\hat{1},2}}\mathrm{d}\bar{\eta}_{n,1}^{x_{n-\hat{1},1}}\mathrm{d}\bar{\xi}_{n^{*}}^{t_{n^{*}-\hat{2},\mathrm{f}}}\nonumber\\
  &\cdot\left(\bar{\eta}_{n+\hat{1}+\hat{1},1}\eta_{n+\hat{1},1}\right)^{x_{n+\hat{1},1}}\left(\bar{\eta}_{n+\hat{1},2}\eta_{n+\hat{1}+\hat{1},2}\right)^{x_{n+\hat{1},2}}\left(\bar{\xi}_{n^{*}+\hat{2}}\xi_{n^{*}}\right)^{t_{n^{*}},\mathrm{f}}\nonumber\\
  &\cdot\delta_{(t_{n,1}+t_{n,2}+t_{n+\hat{1},1}+t_{n+\hat{1},2})\bmod 2,t_{n^{*},\mathrm{f}}}, \nonumber
\end{align}
where the hopping factors for the new Grassmann variable have been shifted $n^*\rightarrow n^*+\hat{2}$.
Note that new constraints including the new indices hold:
\begin{align}
  \left(\sum_{i}x_{n+\hat{1},i}+t_{n^{*},\mathrm{f}}+\sum_{i}x_{n-\hat{1},i}+t_{n^{*}-\hat{2},\mathrm{f}}\right)\bmod 2=0,&&\text{for all }n\label{eq:newCondition1}.
\end{align}
This constraint is a consequence of Eq.~(\ref{eq:condition1}) and Eq.~(\ref{eq:condition2}).
From a practical point of view, it is convenient to explicitly multiply the factor
\begin{align}
  \delta_{(\sum_{i}x_{n+\hat{1},i}+t_{n^{*},\mathrm{f}}+\sum_{i}x_{n-\hat{1},i}+t_{n^{*}-\hat{2},\mathrm{f}})\bmod 2,0}
  \label{eq:delta}
\end{align}
to the new tensor.

As mentioned around Eq.~(\ref{eq:13}), the sign factor in Eq.~(\ref{eq:BeforeRename}) should be incorporated into the new bosonic tensor (before renaming the indices),
\begin{align}
  \label{eq:TnewGrassmann0}
  T^{\mathrm{new,b}}_{x_{n+\hat{1}}t_{n^{*},\mathrm{b}}x_{n-\hat{1}}t_{n^{*}-\hat{2},\mathrm{b}}}
  =\sum_{t^{+}_{n},t^{-}_{n},x_{n}}^{\mathrm{all}}
  \begin{aligned}[t]
    &U^{*}_{t^{+}_{n}t_{n^{*},\mathrm{b}}}
    T_{x_{n}t_{n}x_{n-\hat{1}}t_{n-\hat{2}}}T_{x_{n+\hat{1}}t_{n+\hat{1}}x_{n}t_{n+\hat{1}-\hat{2}}}
    U_{t^{-}_{n}t_{n^{*}-\hat{2},\mathrm{b}}} \\
    &\cdot\left(-1\right)^{x_{n,1}\left(x_{n,1}+x_{n,2}\right)+t_{n,2}\left(t_{n,1}+t_{n,2}\right)+t_{n+\hat{1},2}\left(t_{n+\hat{1},1}+t_{n+\hat{1},2}\right)}.
  \end{aligned}
\end{align}
One can include the phase factors when contracting $M$ and $U$ consist of purely bosonic $T$ in Eq.~(\ref{eq:MakeM}) or
include them when making $M$.
We observe that the latter shows a slightly better approximation.
In Sec.~\ref{sec:RESULTS}, we show the results in the latter case.
For later use, we summarize the new bosonic tensor including the constraint in Eq.~(\ref{eq:delta}),
\begin{align}
  \label{eq:TnewGrassmann}
  T^{\mathrm{new}}_{x_{n+\hat{1}}t_{n^{*}}x_{n-\hat{1}}t_{n^{*}-\hat{2}}}
  =&
     T_{x_{n}t_{n,\mathrm{b}}x_{n-\hat{1}^{*}}t_{n-\hat{2},\mathrm{b}}}^{\mathrm{new,b}}
     \delta_{(\sum_{i}x_{n+\hat{1},i}+t_{n^{*},\mathrm{f}}+\sum_{i}x_{n-\hat{1},i}+t_{n^{*}-\hat{2},\mathrm{f}})\bmod 2,0},
\end{align}
where the index $t$ in the LHS should be understood as
$t_{n^{*}}=(t_{n^{*},\mathrm{f}},t_{n^{*},\mathrm{b}})$ and
$t_{n^{*}-\hat2}=(t_{n^{*}-\hat2,\mathrm{f}},t_{n^{*}-\hat2,\mathrm{b}})$
while $x_{n}$ and $x_{n-\hat{1}^{*}}$ are the same as before in Eq.~(\ref{eq:x12}).

Following the same rule in Eq.~(\ref{eq:Rename}) for the bosonic part,
the Grassmann part of the new tensor in Eq.~(\ref{eq:BeforeRename}) is renamed as follows,
\begin{align}
  &\mathrm{d}\bar{\eta}_{n+\hat{1},2}^{x_{n+\hat{1},2}}\mathrm{d}\eta_{n+\hat{1},1}^{x_{n+\hat{1},1}}\mathrm{d}\xi_{n^{*}}^{t_{n^{*},\mathrm{f}}}\mathrm{d}\eta_{n,2}^{x_{n-\hat{1},2}}\mathrm{d}\bar{\eta}_{n,1}^{x_{n-\hat{1},1}}\mathrm{d}\bar{\xi}_{n^{*}}^{t_{n^{*}-\hat{2},\mathrm{f}}}\nonumber\\
  &\cdot\left(\bar{\eta}_{n+\hat{1}+\hat{1},1}\eta_{n+\hat{1},1}\right)^{x_{n+\hat{1},1}}\left(\bar{\eta}_{n+\hat{1},2}\eta_{n+\hat{1}+\hat{1},2}\right)^{x_{n+\hat{1},2}}\left(\bar{\xi}_{n^{*}+\hat{2}}\xi_{n^{*}}\right)^{t_{n^{*}},\mathrm{f}}\nonumber\\
  &\rightarrow
    \begin{aligned}[t]
      &\mathrm{d}\bar{\eta}_{n^{*},2}^{x_{n^{*},2}}\mathrm{d}\eta_{n^{*},1}^{x_{n^{*},1}}\mathrm{d}\xi_{n^{*}}^{t_{n^{*},\mathrm{f}}}\mathrm{d}\eta_{n^{*},2}^{x_{n^{*}-\hat{1}^{*},2}}\mathrm{d}\bar{\eta}_{n^{*},1}^{x_{n^{*}-\hat{1}^{*},1}}\mathrm{d}\bar{\xi}_{n^{*}}^{t_{n^{*}-\hat{2},\mathrm{f}}}\\
      &\cdot\left(\bar{\eta}_{n^{*}+\hat{1}^{*},1}\eta_{n^{*},1}\right)^{x_{n^{*},1}}\left(\bar{\eta}_{n^{*},2}\eta_{n^{*}+\hat{1}^{*},2}\right)^{x_{n^{*},2}}\left(\bar{\xi}_{n^{*}+\hat{2}}\xi_{n^{*}}\right)^{t_{n^{*}},\mathrm{f}}
    \end{aligned}\nonumber\\
  &\rightarrow
    \begin{aligned}[t]
      &\mathrm{d}\bar{\eta}_{n,2}^{x_{n,2}}\mathrm{d}\eta_{n,1}^{x_{n,1}}\mathrm{d}\xi_{n}^{t_{n,\mathrm{f}}}\mathrm{d}\eta_{n,2}^{x_{n-\hat{1}^{*},2}}\mathrm{d}\bar{\eta}_{n,1}^{x_{n-\hat{1}^{*},1}}\mathrm{d}\bar{\xi}_{n}^{t_{n-\hat{2},\mathrm{f}}}\\
      &\cdot\left(\bar{\eta}_{n+\hat{1}^{*},1}\eta_{n,1}\right)^{x_{n,1}}\left(\bar{\eta}_{n,2}\eta_{n+\hat{1}^{*},2}\right)^{x_{n,2}}\left(\bar{\xi}_{n+\hat{2}}\xi_{n}\right)^{t_{n},\mathrm{f}}.
    \end{aligned}
\end{align}
The ordering of the Grassmann variables and the index assignment
are consistent with that of the initial tensor in Eq.~(\ref{eq:FirstT}).

By combining the coarse-graining of the bosonic part and the Grassmann part along the $\hat{1}$-direction,
we obtain a coarse-grained tensor after the renaming
\begin{align}
  \mathcal{T}_{x_{n}t_{n}x_{n-\hat{1}^{*}}t_{n-\hat{2}}}=
  \begin{aligned}[t]
    &T_{x_{n}t_{n}x_{n-\hat{1}^{*}}t_{n-\hat{2}}}^{\mathrm{new}}\mathrm{d}\bar{\eta}_{n,2}^{x_{n,2}}\mathrm{d}\eta_{n,1}^{x_{n,1}}\mathrm{d}\xi_{n}^{t_{n,\mathrm{f}}}\mathrm{d}\eta_{n,2}^{x_{n-\hat{1}^{*},2}}\mathrm{d}\bar{\eta}_{n,1}^{x_{n-\hat{1}^{*},1}}\mathrm{d}\bar{\xi}_{n}^{t_{n-\hat{2},\mathrm{f}}}\\
    &\cdot\left(\bar{\eta}_{n+\hat{1}^{*},1}\eta_{n,1}\right)^{x_{n,1}}\left(\bar{\eta}_{n,2}\eta_{n+\hat{1}^{*},2}\right)^{x_{n,2}}\left(\bar{\xi}_{n+\hat{2}}\xi_{n}\right)^{t_{n},\mathrm{f}}.
  \end{aligned}
      \label{0.5T}
\end{align}
The Grassmann part of the above form is similar to that of the initial tensor in Eq.~(\ref{eq:FirstT}).
Actually, by replacing
$t_{n,1}\rightarrow t_{n,\mathrm{f}}$,
$t_{n,2}\rightarrow0$,
and
$\xi_{n,1}\rightarrow\xi_{n}$ (a single component)
in Eq.~(\ref{eq:FirstT}),
one obtains the above structure.

The second half of the coarse-graining along the $\hat{2}$-direction can be done as follows.
First, let us interchange the $\hat{1}$- and the $\hat{2}$-axis and obtain a flipped tensor
\begin{align}
  \label{eq:6}
  \mathcal{T}_{t_{n}x_{n}t_{n-\hat{2}}x_{n-\hat{1}^{*}}}=\left(-1\right)^{\left(x_{n,\mathrm{1}}+x_{n,\mathrm{2}}\right)t_{n,\mathrm{f}}+\left(x_{n-\hat{1}^{*},1}+x_{n-\hat{1}^{*},2}\right)t_{n-\hat{2},\mathrm{f}}}\mathcal{T}_{x_{n}t_{n}x_{n-\hat{1}^{*}}t_{n-\hat{2}}}.
\end{align}
By applying a similar coarse-graining procedure for the $\hat{1}$-direction to this tensor,
one can carry out the coarse-graining for the $\hat{2}$-direction.
The coarse-grained tensor is then obtained by
\begin{align}
  \label{eq:1T}
  \mathcal{T}_{x_{n}t_{n}x_{n-\hat{1}^{*}}t_{n-\hat{2}^{*}}}=T_{x_{n}t_{n}x_{n-\hat{1}^{*}}t_{n-\hat{2}^{*}}}\mathrm{d}\eta_{n}^{x_{n,\mathrm{f}}}\mathrm{d}\xi_{n}^{t_{n,\mathrm{f}}}\mathrm{d}\bar{\eta}_{n}^{x_{n-{\hat{1}^{*}}}}\mathrm{d}\bar{\xi}_{n}^{t_{n-\hat{2}^{*},\mathrm{f}}}\left(\bar{\eta}_{n+\hat{1}^{*}}\eta_{n}\right)^{x_{n,\mathrm{f}}}\left(\bar{\xi}_{n+\hat{2}^{*}}\xi_{n}\right)^{t_{n,\mathrm{f}}},
\end{align}
where the all indices of the tensor have the structure $x_{n}=(x_{n,\mathrm{f}},x_{n,\mathrm{b}})$.
The sign factor and the constraint have been already included in the bosonic tensor.
Now, the scale of the $\hat{1}$-direction is equal to that of the $\hat{2}$-direction.
We define an ``iteration'' of coarse-graining which consists of the coarse-graining for both directions.
Then Eq.~(\ref{eq:1T}) is recognized as the coarse-grained tensor after the first iteration as well as
the starting tensor of the second iteration.
Actually, it turns out that from the second iteration, the structure of tensor in Eq.~(\ref{eq:1T}) does not change.
The actual coarse-graining procedure from the second iteration can be done by replacing the second components of the indices to zero in the first iteration.
For example, for the $\hat1$-direction one only has to replace
\begin{align}
  x_{n,2}\rightarrow0, \quad t_{n,2}\rightarrow0,&&\text{for all }n
\end{align}
in the Grassmann part.

The GHOTRG algorithm presented here can be immediately extended to any dimensional system.
We do not show any details of a coarse-graining for higher dimensional systems,
but we present an initial tensor for the three dimensional free Wilson fermion with two spinor components in App.~\ref{sec:TNRepresentation}.

\subsection{Anti-periodic boundary conditions}
\label{sec:APBC}

\begin{figure}[htbp]
  \centering
  \includegraphics{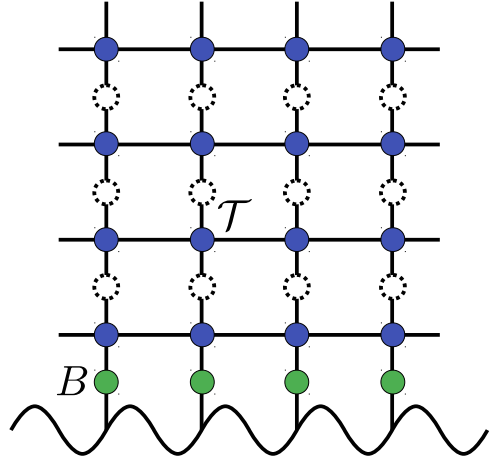}
  \caption{Insertion of tensor $B$ in the tensor network. This insertion only affects the boundary of the network while
    the bulk is intact.}
  \label{insert_B}
\end{figure}

If anti-periodic boundary conditions are imposed, the sign factor $\left(-1\right)$ arises when a fermion line passes the boundary for the $\hat{2}$-direction.
Therefore, one can realize anti-periodic boundary conditions on the tensor network by inserting a boundary tensor $B$ 
into each link along the $\hat{2}$-direction as shown in Fig.~\ref{insert_B},
\begin{align}
    &\mathcal{T}_{x_{n}t_{n}x_{n-\hat{1}}t_{n-\hat{2}}}\rightarrow \mathcal{T}_{x_{n}t_{n}x_{n-\hat{1}}t'_{n}}B_{t'_{n}t_{n-\hat{2}}},\\
    &B_{t'_{n}t_{n-\hat{2}}}=
      \begin{cases}
        \left(-1\right)^{t'_{n,1}+t'_{n,2}}\delta_{t'_{n},t_{n-\hat{2}}}, & \text{if }n_2=0,\\
        \delta_{t'_{n},t_{n-\hat{2}}}, & \text{otherwise.}
      \end{cases}
\end{align}
This insertion affects only the tensors around the boundary.

\begin{figure}[htbp]
  \centering
  \includegraphics{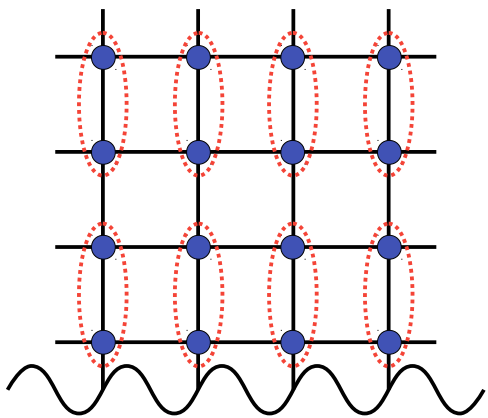}
  \caption{Coarse-graining along the $\hat{2}$-direction in the presence of the boundary. There is no difficulty for the coarse-graining as long as the number of sites for the $\hat 2$-direction is even.}
  \label{trivial}
\end{figure}

\begin{figure}[htbp]
  \centering
  \includegraphics{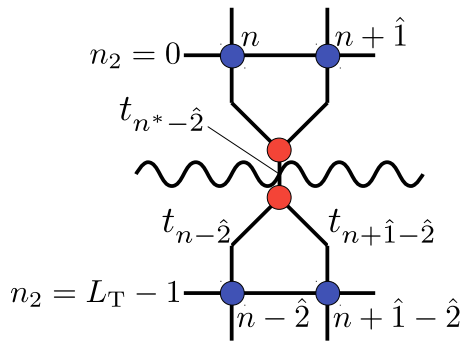}
  \caption{Coarse-graining along the $\hat{1}$-direction. For the coarse-graining involving the boundary, the additional sign factors
    originated from boundary tensors should be taken care.}
  \label{boundary}
\end{figure}

Even when there is a boundary, the coarse-graining along the $\hat{2}$-direction is simple.
As shown in Fig.~\ref{trivial}, 
there is no difficulty for the coarse-graining as long as the number of lattice sites for this direction is even.
The coarse-graining along the $\hat{1}$-direction is also straightforward although one needs to care about the additional sign factor in $B$.
In the coarse-graining step involving the boundary shown in Fig.~\ref{boundary},
as a result of the insertion of $B$, a sign factor
\begin{align}
  \label{B}
  \left(-1\right)^{t_{n-\hat{2},1}+t_{n-\hat{2},2}+t_{n+\hat{1}-\hat{2},1}+t_{n+\hat{1}-\hat{2},2}}
\end{align}
appears.
By using the new index $t_{n^{*}-\hat{2},\mathrm{f}}$ in Eq.~(\ref{eq:condition2}), the factor is rewritten as
\begin{align}
  \left(-1\right)^{t_{n-\hat{2},1}+t_{n-\hat{2},2}+t_{n+\hat{1}-\hat{2},1}+t_{n+\hat{1}-\hat{2},2}}&=\left(-1\right)^{\left(t_{n-\hat{2},1}+t_{n-\hat{2},2}+t_{n+\hat{1}-\hat{2},1}+t_{n+\hat{1}-\hat{2},2}\right)\bmod 2}\nonumber\\
                                                                                                   &=\left(-1\right)^{t_{n^{*}-\hat{2},\mathrm{f}}}.
\end{align}
Therefore we define a new boundary tensor $B^{\mathrm{new}}$ for the next coarse-graining,
\begin{align}
                                      B^{\mathrm{new}}_{t'_{n}t_{n-\hat{2}}}=
                                      \begin{cases}
                                        \left(-1\right)^{t'_{n,\mathrm{f}}}\delta_{t'_{n},t_{n-\hat{2}}}, & \text{if }n_2=0,\\
                                        \delta_{t'_{n},t_{n-\hat{2}}}, & \text{otherwise}.
                                      \end{cases}
\end{align}
From the second iteration, the structure of boundary tensor does not change.
In this way, the boundary effect is not involved in the coarse-graining procedure for the tensor ${\cal T}$ in the bulk
and appears only in the contraction procedure to compute the partition function.

\section{Numerical results}
\label{sec:RESULTS}
\subsection{Two dimensional systems}
\subsubsection{The error of the free energy and the hierarchy of eigenvalues}

\begin{figure}[htbp]
  \centering
  \includegraphics{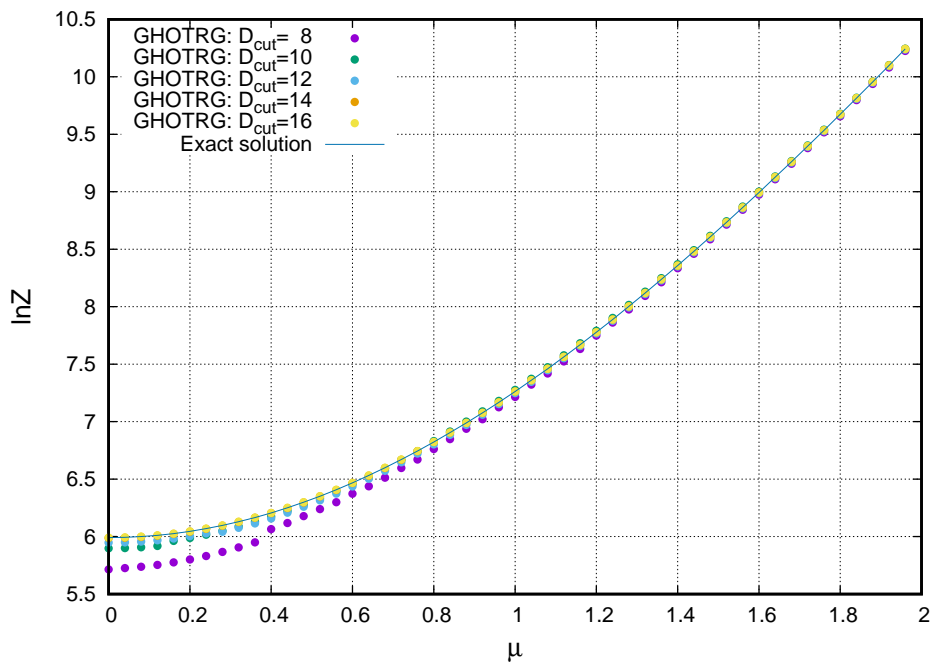}
  \caption{The free energy of massless free Wilson fermions as a function of $\mu$ on $2\times 2$ lattice.}
  \label{lnZ}
\end{figure}

Here we exclusively consider the massless free Wilson fermion.
Figure~\ref{lnZ} shows the free energy as a function of the chemical potential $\mu$ on the $2\times 2$ space-time lattice computed by using the GHOTRG.
For the maximal $D_{\mathrm{cut}}\left(=16\right)$, it agrees with the analytical exact result up to the machine precision.
On the other hand, when $D_{\mathrm{cut}}$ decreases, the difference becomes larger especially in the region $\mu <1.0$.

\begin{figure}
  \centering
  \includegraphics{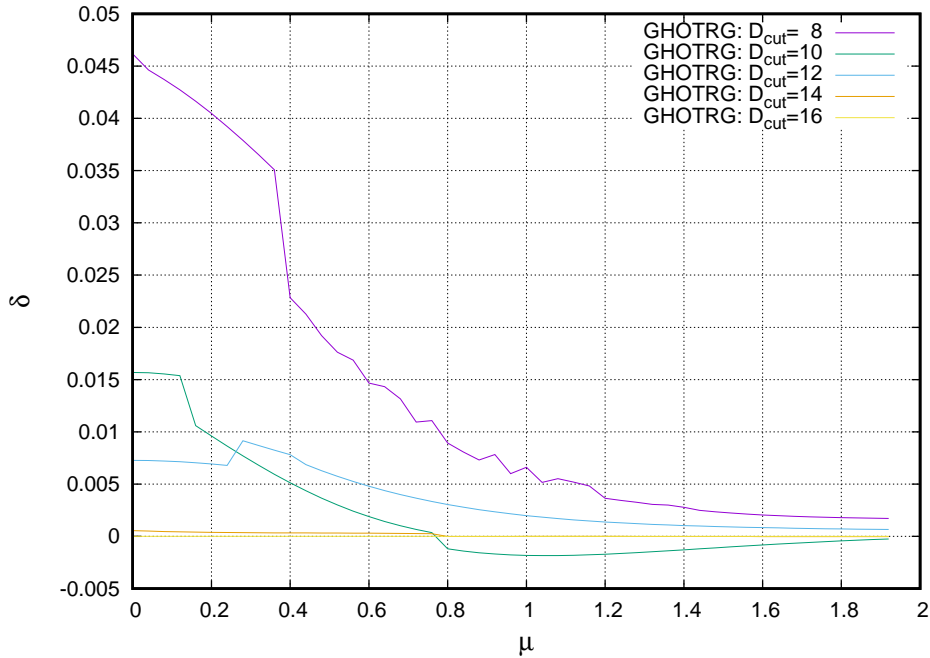}
  \caption{The error of the free energy for massless free Wilson fermions as a function of $\mu$ on $2\times 2$ lattice.}
  \label{diff}
\end{figure}

\begin{figure}[htbp]
  \centering
  \includegraphics{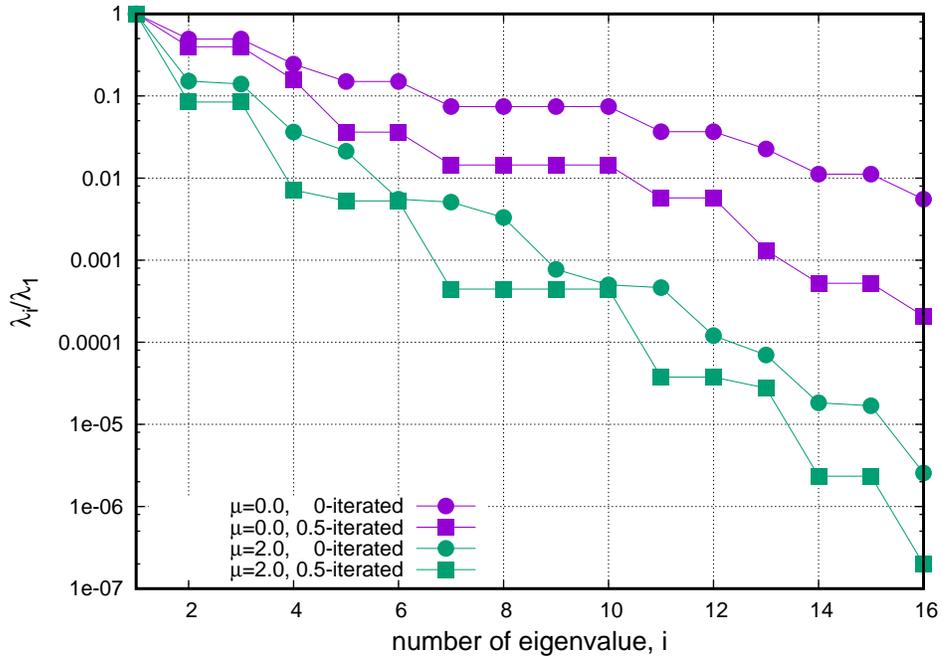}
  \caption{The hierarchy of eigenvalues of $M^{\pm}$ at $\mu=0.0$ and $\mu=2.0$ with $D_{\mathrm{cut}}=16$ for massless free system. An anisotropic coarse-graining along the $\hat{1}$-direction is called 0.5-iteration.}
  \label{fig:EV_Dcut16_mu0.0-2.0}
\end{figure}

Figure~\ref{diff} shows the error defined as
\begin{align}
  \label{eq:25}
  \delta=\frac{\ln{Z_{\mathrm{exact}}}-\ln{Z\left(D_{\mathrm{cut}}\right)}}{\left|\ln{Z_{\mathrm{exact}}}\right|}.
\end{align}
For larger $D_{\mathrm{cut}}$, the error rapidly reduces as expected. 
In the figure, we observe that the error becomes larger for $\mu\approx 0.0$ while it is smaller for $\mu\approx 2.0$.
This tendency is seen for all $D_{\mathrm{cut}}<16$.
In order to qualitatively understand this behavior, we investigate the hierarchy of eigenvalues of $M^{\pm}$~\footnote{Note that the phase factors which have arisen in the Grassmann part are included in $M^{\pm}$.}.
We show them at $\mu=0.0$ and $\mu=2.0$ in Fig.~\ref{fig:EV_Dcut16_mu0.0-2.0}.
For $\mu=0.0$, the hierarchy is milder than that of $\mu=2.0$.
Therefore we confirm the strong correlation between the error of the free energy and the hierarchy of eigenvalues;
a low compression in the coarse-graining procedure causes a large error in a physical quantity.

\begin{figure}
  \centering
  \includegraphics{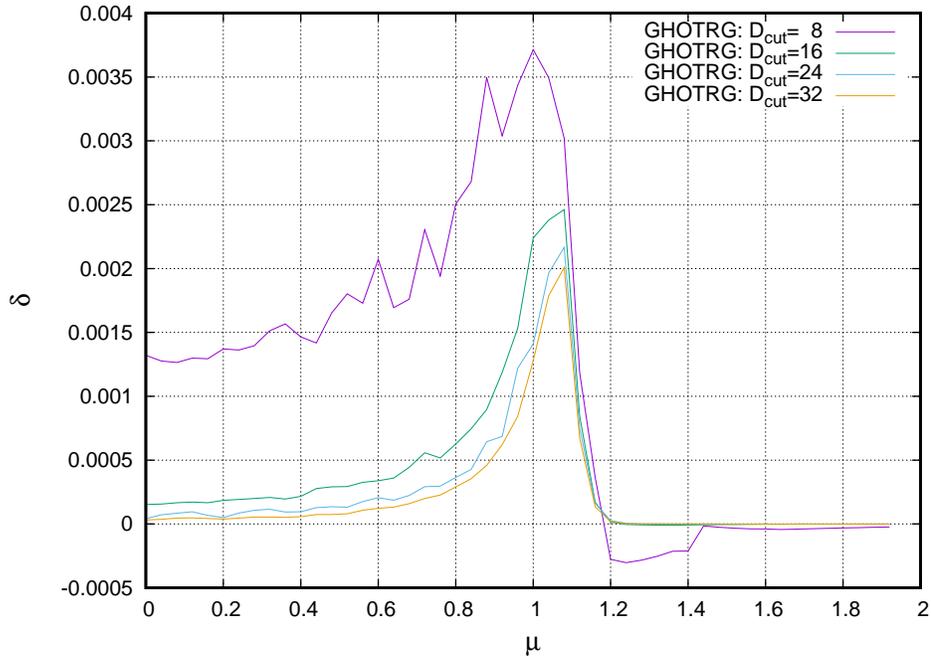}
  \caption{The error of the free energy for massless free Wilson fermions as a function of $\mu$ on $32\times 32$ lattice.}
  \label{fig:L32_diff8-32}
\end{figure}

\begin{figure}[htbp]
  \centering
  \includegraphics{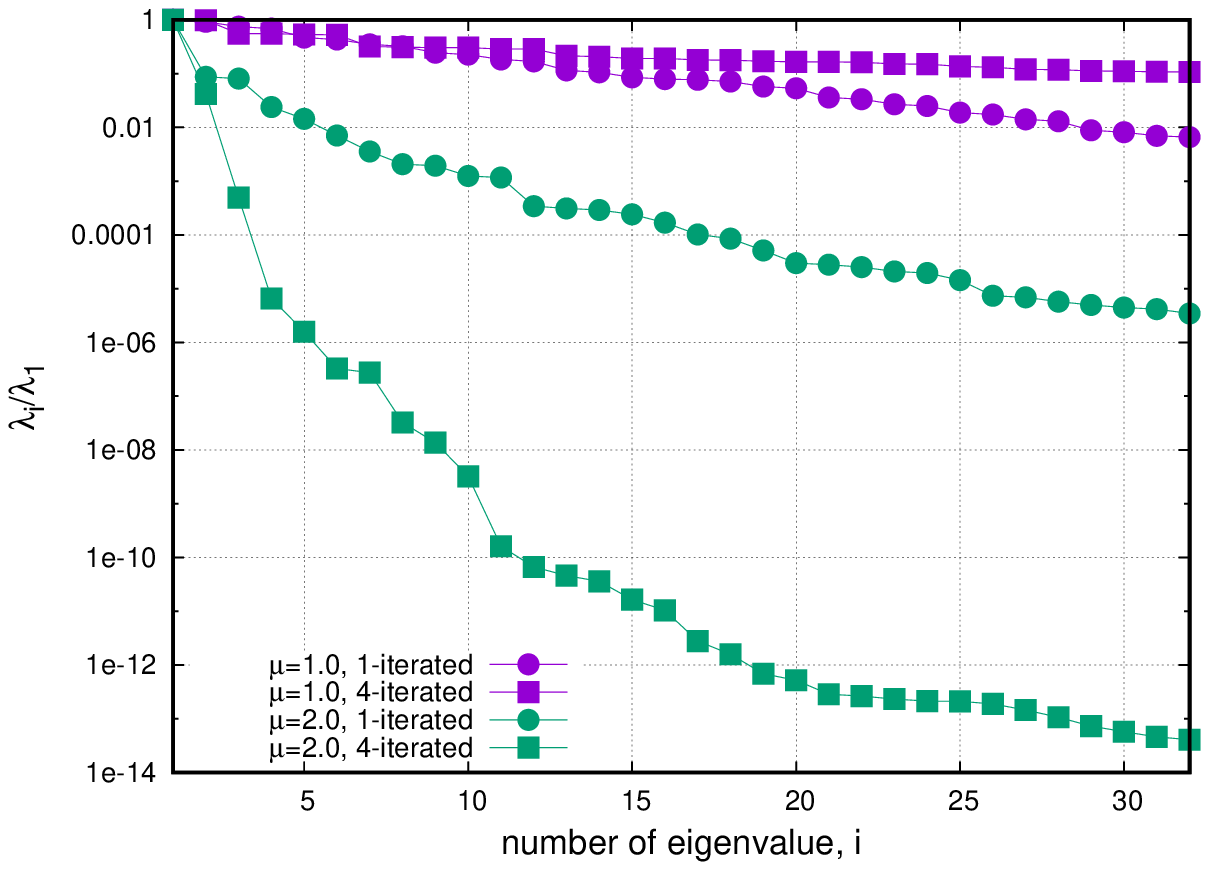}
  \caption{The hierarchy of eigenvalues at $\mu=1.0$ and $\mu=2.0$ with $D_{\mathrm{cut}}=32$ for the massless free system.}
  \label{fig:EV_Dcut32_mu1.0-2.0}
\end{figure}

We also investigate the error for a larger lattice volume $32\times 32$ as shown in Fig.~\ref{fig:L32_diff8-32}.
The error tends to be large around $\mu=1.0$.
The eigenvalues of $M^{\pm}$ is shown in Fig.~\ref{fig:EV_Dcut32_mu1.0-2.0} where
the hierarchy at $\mu=1.0$ is quite poor compared with that of $\mu=2.0$.
This behavior is very similar to that of GTRG~\cite{Takeda:2014vwa}.

\subsubsection{The fermion number density of the lattice Thirring model}

\begin{figure}[htbp]
  \centering
  \includegraphics{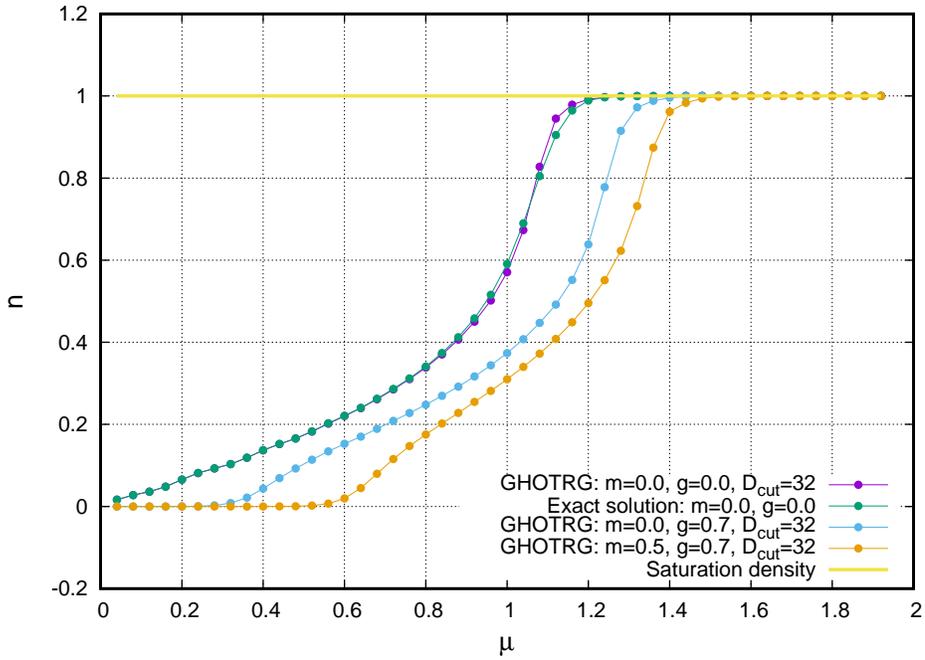}
  \caption{The fermion number density for the lattice Thirring model. The exact results for massless free case is also shown. The lattice volume is $32\times 32$.}
  \label{fig:NumberDensity32_g0.7}
\end{figure}

Figure~\ref{fig:NumberDensity32_g0.7} shows the fermion number density
\begin{align}
  n=\frac{1}{V}\frac{\partial{\ln{Z}}}{\partial \mu}
\end{align}
for the lattice Thirring model whose tensor network representation is presented in Ref.~\cite{Takeda:2014vwa}.
In the large $\mu$ region, the fermion number density reaches the saturation density.
When switching on the interaction $g\neq0$, the Silver Blaze like phenomenon seems to occur around the small $\mu$ region where the fermion number density is constantly zero while the onset is observed at finite $\mu$.
Such a behavior was already seen in Ref.~\cite{Takeda:2014vwa} where GTRG is used.
In any case, the exact massless free results are well reproduced by GHOTRG with $D_{\rm cut}=32$ in a wide range of $\mu$
although the approximation gets worse around $\mu=1.0$; this tendency was also observed in the GTRG analysis~\cite{Takeda:2014vwa}.

\subsection{Three dimensional system}
\subsubsection{The free energy}

\begin{figure}[htbp]
  \centering
  \includegraphics{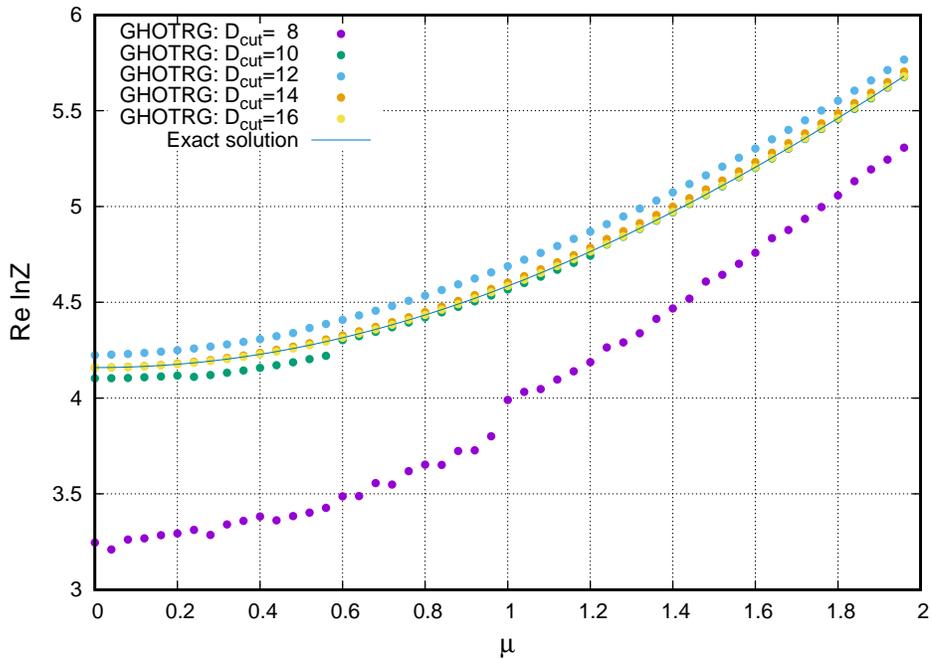}
  \caption{The free energy of three dimensional massless free Wilson fermions as a function of $\mu$ on $2\times 1\times 1$ lattice.
    The real part of the free energy is shown.}
  \label{fig:L211_Dcut8-16}
\end{figure}

We apply GHOTRG to the three dimensional massless free Wilson fermion and calculate the free energy.
Since it is very demanding to carry out the full $D_{\rm cut}$ calculation even for the $2\times2\times2$ lattice,
we have to compromise to use an anisotropic lattice, say, $2\times1\times1$ to check the algorithm and the code.
As a check, we carry out the coarse-graining from $2\times 1\times 1$ to $1\times 1\times 1$ with several $D_{\mathrm{cut}}$
as shown in Fig.~\ref{fig:L211_Dcut8-16}.
With the full $D_{\mathrm{cut}}=16$, the GHOTRG result is consistent with the exact result up to the machine precision.
Even for smaller $D_{\rm cut}<16$, the accuracy is reasonable, and for larger $D_{\rm cut}$ the free energy rapidly converges to the exact one.
Note that at the smaller $D_{\rm cut}$, the free energy may be complex due to the truncation in the coarse-graining procedure;
thus the real part of the free energy is shown in the figure.

\section{Summary and outlook}
\label{sec:SUMMARY}

In this paper we have formulated Grassmann higher order tensor renormalization group and applied it to the free fermion system and the lattice Thirring model.
The numerical results are consistent with analytical or previous ones. Thus we conclude that GHOTRG is a correct algorithm.

In the study of two dimensional finite chemical potential systems, we observed the very poor hierarchy of eigenvalues at $\mu=1.0$,
and this is the reason why the accuracy gets worse around this parameter region.
This shows that the situation of accuracy for GHOTRG is similar to that of the GTRG.

The algorithm of GHOTRG can be straightforwardly extended to three dimensions or more.
In principle, now one can deal with any dimensional fermion system by using GHOTRG.
It is, however, still hard to study four dimensional fermion systems using GHOTRG owing to a huge computational cost.
We hope that our formulation is a starting point towards the four dimensional lattice QCD.
Apart from lattice QCD, there are some interesting models in low dimensions such as lattice chiral gauge theories and lattice SUSY which are generally suffering from the sign problem.
We are now in the position to approach such models in two or three dimensions by using GHOTRG.

The bottleneck of (G)HOTRG is a tensor contraction whose cost is proportional to $D_{\rm cut}^{4d-1}$ with the dimensionality $d$.
This cost could be reduced by randomizing the contraction.
So far, many random algorithms have been proposed in the frameworks of MPS, PEPS, and TTN~\cite{2007PhRvL..99v0602S,schuch2008simulation,2009NJPh...11h3026M}.
Recently, new algorithms combined with MERA are proposed~\cite{2012PhRvB..85p5146F,2012PhRvB..85p5147F}.
A tensor network Monte Carlo method introduced by Ferris~\cite{2015arXiv150700767F} may be applied to the HOTRG.
It seems worthwhile to pursue this direction.

\section*{Acknowledgments}
We acknowledge helpful discussions with H. Kawauchi. 
R. S. would like to thank Prof. Y. Kuramashi for giving an opportunity to stay at RIKEN AICS.
R. S. also acknowledge valuable discussions with Dr. Y. Nakamura and Dr. Y. Shimizu.
This work is supported by JSPS KAKENHI Grant Numbers 17K05411, Kanazawa University SAKIGAKE Project, and MEXT as ``Exploratory Challenge on Post-K computer'' (Frontiers of Basic Science: Challenging the Limits).


\appendix

\section{Tensor network representation of the free fermion system in three dimensions}
\label{sec:TNRepresentation}

We explain how to obtain a tensor network representation for the free Wilson fermions with two spinor components on the three dimensional lattice.
The content of this appendix follows the strategy in Ref.~\cite{Takeda:2014vwa}.

By using an explicit representation of the $\gamma$ matrices
\begin{align}
  \gamma_1=\sigma_1=
  \begin{pmatrix}
    0 & 1 \\
    1 & 0
  \end{pmatrix}
        ,\quad
        \gamma_2=\sigma_3=
        \begin{pmatrix}
          1 & 0 \\
          0 & -1
        \end{pmatrix}
              ,\quad
              \gamma_{3}=\sigma_{2}=
              \begin{pmatrix}
                0 & -i \\
                i & 0
              \end{pmatrix}
                    ,
\end{align}
one can write down the lattice action as
\begin{align}
  S_{\mathrm{F}}=\sum_{n}
  \begin{aligned}[t]
    \Biggl[&\left(m+3\right)\bar{\psi}_{n}\psi_{n}-\frac{1}{2}\bar{\psi}_{n}
    \begin{pmatrix}
      1 & 1\\
      1 & 1
    \end{pmatrix}
    \psi_{n-\hat{1}}-\frac{1}{2}\bar{\psi}_{n}
    \begin{pmatrix}
      1 & -1\\
      -1 & 1
    \end{pmatrix}
    \psi_{n+\hat{1}}\\
    &-e^{-\mu}\bar{\psi}_{n}
    \begin{pmatrix}
      1 & 0\\
      0 & 0
    \end{pmatrix}
    \psi_{n-\hat{2}}-e^{\mu}\bar{\psi}_{n}
    \begin{pmatrix}
      0 & 0\\
      0 & 1
    \end{pmatrix}
    \psi_{n+\hat{2}}\\
    &-\frac{1}{2}\bar{\psi}_{n}
    \begin{pmatrix}
      1 & -i \\
      i & 1
    \end{pmatrix}
    \psi_{n-\hat{3}}-\frac{1}{2}\bar{\psi}_{n}
    \begin{pmatrix}
      1 & i \\
      -i & 1
    \end{pmatrix}
    \psi_{n+\hat{3}}\Biggr].
  \end{aligned}
\end{align}
As you can see, the hopping terms along the $\hat{1}$-direction and the $\hat{3}$-direction mix the different spinor components.
Here, we introduce new linear combinations as
\begin{align}
  &\chi_{n,1}=\frac{1}{\sqrt{2}}\left(\psi_{n,1}+\psi_{n,2}\right),\quad \chi_{n,2}=\frac{1}{\sqrt{2}}\left(\psi_{n,1}-\psi_{n,2}\right),\\
  &\bar{\chi}_{n,1}=\frac{1}{\sqrt{2}}\left(\bar{\psi}_{n,1}+\bar{\psi}_{n,2}\right),\quad \bar{\chi}_{n,2}=\frac{1}{\sqrt{2}}\left(\bar{\psi}_{n,1}-\bar{\psi}_{n,2}\right),\\
  &\phi_{n,1}=\frac{1}{\sqrt{2}}\left(\psi_{n,1}-i\psi_{n,2}\right),\quad \phi_{n,2}=\frac{1}{\sqrt{2}}\left(\psi_{n,1}+i\psi_{n,2}\right),\\
  &\bar{\phi}_{n,1}=\frac{1}{\sqrt{2}}\left(\bar{\psi}_{n,1}+i\bar{\psi}_{n,2}\right),\quad \bar{\phi}_{n,2}=\frac{1}{\sqrt{2}}\left(\bar{\psi}_{n,1}-i\bar{\psi}_{n,2}\right).
\end{align}
By using the new fields, one obtains a diagonal representation in the spinor space
\begin{align}
  S_{\mathrm{F}}&=
                  \begin{aligned}[t]
                    \sum_{n}\Bigl[&\left(m+3\right)\bar{\psi}_{n,1}\psi_{n,1}+\left(m+3\right)\bar{\psi}_{n,2}\psi_{n,2}-\bar{\chi}_{n,1}\chi_{n-\hat{1},1}-\bar{\chi}_{n,2}\chi_{n+\hat{1},2}\\
                    &-e^{-\mu}\bar{\psi}_{n,1}\psi_{n-\hat{2},1}-e^{\mu}\bar{\psi}_{n,2}\psi_{n+\hat{2},2}-\bar{\phi}_{n,1}\phi_{n-\hat{3},1}-\bar{\phi}_{n,2}\phi_{n+\hat{3},2}\Bigr].
                  \end{aligned}
\end{align}

Thanks to the nilpotency of Grassmann variables, each exponential factor in the partition function can be expanded binomially:
\begin{align}
  Z&=\int\mathcal{D}\psi\mathcal{D}\bar{\psi}\prod_{n}
     \begin{aligned}[t]
       &e^{-\left(m+3\right)\bar{\psi}_{n,1}\psi_{n,1}}e^{-\left(m+3\right)\bar{\psi}_{n,2}\psi_{n,2}}e^{\bar{\chi}_{n,1}\chi_{n-\hat{1},1}}e^{\bar{\chi}_{n,2}\chi_{n+\hat{1},2}}\\
       &\cdot e^{e^{-\mu}\bar{\psi}_{n,1}\psi_{n-\hat{2},1}}e^{e^{\mu}\bar{\psi}_{n,2}\psi_{n+\hat{2},2}}e^{\bar{\phi}_{n,1}\phi_{n-\hat{3},1}}e^{\bar{\phi}_{n,2}\phi_{n+\hat{3},2}}
     \end{aligned}\nonumber\\
   &=\sum_{\left\{a,x,t,y\right\}}\int\mathcal{D}\psi\mathcal{D}\bar{\psi}\prod_{n}
     \begin{aligned}[t]
       &\left(-\left(m+3\right)\bar{\psi}_{n,1}\psi_{n,1}\right)^{a_{n,1}}\left(-\left(m+3\right)\bar{\psi}_{n,2}\psi_{n,2}\right)^{a_{n,2}}\\
       &\cdot \left(\bar{\chi}_{n,1}\chi_{n-\hat{1},1}\right)^{x_{n,1}}\left(\bar{\chi}_{n,2}\chi_{n+\hat{1},2}\right)^{x_{n,2}}\left(e^{-\mu}\bar{\psi}_{n,1}\psi_{n-\hat{2},1}\right)^{t_{n,1}}\\
       &\cdot \left(e^{\mu}\bar{\psi}_{n,2}\psi_{n+\hat{2},2}\right)^{t_{n,2}}\left(\bar{\phi}_{n,1}\phi_{n-\hat{3},1}\right)^{y_{n,1}}\left(\bar{\phi}_{n,2}\phi_{n+\hat{3},2}\right)^{y_{n,2}}.
     \end{aligned}
\end{align}
In order to integrate out the original fields $\psi_{n}$, $\bar{\psi}_{n}$ (, $\chi_{n}$, $\bar{\chi}_{n}$, $\phi_{n}$, and $\bar{\phi}_{n}$), we insert new Grassmann variables $\eta_{n}$, $\bar{\eta}_{n}$, $\xi_{n}$, $\bar{\xi}_{n}$, $\zeta_{n}$, and $\bar{\zeta}_{n}$ in the hopping factors
\begin{align}
  &\left(\bar{\chi}_{n+\hat{1},1}\chi_{n,1}\right)^{x_{n,1}}=\left(\chi_{n,1}\mathrm{d}\eta_{n,1}\right)^{x_{n,1}}\left(\bar{\chi}_{n+\hat{1},1}\eta_{n,1}\right)^{x_{n,1}},\\
  &\left(\bar{\chi}_{n,2}\chi_{n+\hat{1},2}\right)^{x_{n,2}}=\left(\bar{\chi}_{n,2}\mathrm{d}\bar{\eta}_{n,2}\right)^{x_{n,2}}\left(\bar{\eta}_{n,2}\chi_{n+\hat{1},2}\right)^{x_{n,2}},\\
  &\left(e^{-\mu}\bar{\psi}_{n+\hat{2},1}\psi_{n,1}\right)^{t_{n,1}}=\left(e^{-\frac{\mu}{2}}\psi_{n,1}\mathrm{d}\xi_{n,1}\right)^{t_{n,1}}\left(e^{-\frac{\mu}{2}}\bar{\psi}_{n+\hat{2},1}\xi_{n,1}\right)^{t_{n,1}},\\
  &\left(e^{\mu}\bar{\psi}_{n,2}\psi_{n+\hat{2},2}\right)^{t_{n,2}}=\left(e^{\frac{\mu}{2}}\bar{\psi}_{n,2}\mathrm{d}\bar{\xi}_{n,2}\right)^{t_{n,2}}\left(e^{\frac{\mu}{2}}\bar{\xi}_{n,2}\psi_{n+\hat{2},2}\right)^{t_{n,2}},\\
  &\left(\bar{\phi}_{n+\hat{3},1}\phi_{n,1}\right)^{y_{n,1}}=\left(\phi_{n,1}\mathrm{d}\zeta_{n,1}\right)^{y_{n,1}}\left(\bar{\phi}_{n+\hat{3},1}\zeta_{n,1}\right)^{y_{n,1}},\\
  &\left(\bar{\phi}_{n,2}\phi_{n+\hat{3},2}\right)^{y_{n,2}}=\left(\bar{\phi}_{n,2}\mathrm{d}\bar{\zeta}_{n,2}\right)^{y_{n,2}}\left(\bar{\zeta}_{n,2}\phi_{n+\hat{3},2}\right)^{y_{n,2}},\\
  &\left(\bar{\chi}_{n,1}\chi_{n-\hat{1},1}\right)^{x_{n-\hat{1},1}}=\left(\bar{\chi}_{n,1}\mathrm{d}\bar{\eta}_{n,1}\right)^{x_{n-\hat{1},1}}\left(\bar{\eta}_{n,1}\chi_{n-\hat{1},1}\right)^{x_{n-\hat{1},1}},\\
  &\left(\bar{\chi}_{n-\hat{1},2}\chi_{n,2}\right)^{x_{n-\hat{1},2}}=\left(\chi_{n,2}\mathrm{d}\eta_{n,2}\right)^{x_{n-\hat{1},2}}\left(\bar{\chi}_{n-\hat{1},2}\eta_{n,2}\right)^{x_{n-\hat{1},2}},\\
  &\left(e^{-\mu}\bar{\psi}_{n,1}\psi_{n-\hat{2},1}\right)^{t_{n-\hat{2},1}}=\left(e^{-\frac{\mu}{2}}\bar{\psi}_{n,1}\mathrm{d}\bar{\xi}_{n,1}\right)^{t_{n-\hat{2},1}}\left(e^{-\frac{\mu}{2}}\bar{\xi}_{n,1}\psi_{n-\hat{2},1}\right)^{t_{n-\hat{2},1}},\\
  &\left(e^{\mu}\bar{\psi}_{n-\hat{2},2}\psi_{n,2}\right)^{t_{n-\hat{2},2}}=\left(e^{\frac{\mu}{2}}\psi_{n,2}\mathrm{d}\xi_{n,2}\right)^{t_{n-\hat{2},2}}\left(e^{\frac{\mu}{2}}\bar{\psi}_{n-\hat{2},2}\xi_{n,2}\right)^{t_{n-\hat{2},2}},\\
  &\left(\bar{\phi}_{n,1}\phi_{n-\hat{3},1}\right)^{y_{n-\hat{3},1}}=\left(\bar{\phi}_{n,1}\mathrm{d}\bar{\zeta}_{n,1}\right)^{y_{n-\hat{3},1}}\left(\bar{\zeta}_{n,1}\phi_{n-\hat{3},1}\right)^{y_{n-\hat{3},1}},\\
  &\left(\bar{\phi}_{n-\hat{3},2}\phi_{n,2}\right)^{y_{n-\hat{3},2}}=\left(\phi_{n,2}\mathrm{d}\zeta_{n,2}\right)^{y_{n-\hat{3},2}}\left(\bar{\phi}_{n-\hat{3},2}\zeta_{n,2}\right)^{y_{n-\hat{3},2}}.
\end{align}
Note that these insertions do not generate sign factors.
Now, we can collect all factors which include the original fields at a site $n$, namely $\psi_{n}$, $\bar{\psi}_{n}$, $\chi_{n}$, $\bar{\chi}_{n}$, $\zeta_{n}$, and $\bar{\zeta}_{n}$, without invoking sign factors.
By integrating out the original fields at the site $n$, we define the bosonic tensor $T$ by
\begin{align}
  &\int
    \begin{aligned}[t]
      &\mathrm{d}\psi_{n,1}\mathrm{d}\bar{\psi}_{n,1}\mathrm{d}\psi_{n,2}\mathrm{d}\bar{\psi}_{n,2}\sum_{a_{n,1},a_{n,2}}\left(-\left(m+3\right)\bar{\psi}_{n,1}\psi_{n,1}\right)^{a_{n,1}}\left(-\left(m+3\right)\bar{\psi}_{n,2}\psi_{n,2}\right)^{a_{n,2}}\\
      &\cdot\left(\chi_{n,1}\mathrm{d}\eta_{n,1}\right)^{x_{n,1}}\left(\bar{\chi}_{n,2}\mathrm{d}\bar{\eta}_{n,2}\right)^{x_{n,2}}\left(e^{-\frac{\mu}{2}}\psi_{n,1}\mathrm{d}\xi_{n,1}\right)^{t_{n,1}}\left(e^{\frac{\mu}{2}}\bar{\psi}_{n,2}\mathrm{d}\bar{\xi}_{n,2}\right)^{t_{n,2}}\\
      &\cdot\left(\phi_{n,1}\mathrm{d}\zeta_{n,1}\right)^{y_{n,1}}\left(\bar{\phi}_{n,2}\mathrm{d}\bar{\zeta}_{n,2}\right)^{y_{n,2}}\left(\bar{\chi}_{n,1}\mathrm{d}\bar{\eta}_{n,1}\right)^{x_{n-\hat{1},1}}\left(\chi_{n,2}\mathrm{d}\eta_{n,2}\right)^{x_{n-\hat{1},2}}\\
      &\cdot\left(e^{-\frac{\mu}{2}}\bar{\psi}_{n,1}\mathrm{d}\bar{\xi}_{n,1}\right)^{t_{n-\hat{2},1}}\left(e^{\frac{\mu}{2}}\psi_{n,2}\mathrm{d}\xi_{n,2}\right)^{t_{n-\hat{2},2}}\left(\bar{\phi}_{n,1}\mathrm{d}\bar{\zeta}_{n,1}\right)^{y_{n-\hat{3},1}}\left(\phi_{n,2}\mathrm{d}\zeta_{n,2}\right)^{y_{n-\hat{3},2}}
    \end{aligned}\nonumber\\
  &=
    \begin{aligned}[t]
      &T_{x_{n}t_{n}y_{n}x_{n-\hat{1}}t_{n-\hat{2}}y_{n-\hat{3}}}\\
      &\cdot\mathrm{d}\bar{\eta}_{n,2}^{x_{n,2}}\mathrm{d}\eta_{n,1}^{x_{n,1}}\mathrm{d}\bar{\xi}_{n,2}^{t_{n,2}}\mathrm{d}\xi_{n,1}^{t_{n,1}}\mathrm{d}\bar{\zeta}_{n,2}^{y_{n,2}}\mathrm{d}\zeta_{n,1}^{y_{n,1}}\mathrm{d}\eta_{n,2}^{x_{n-\hat{1},2}}\mathrm{d}\bar{\eta}_{n,1}^{x_{n-\hat{1},1}}\mathrm{d}\xi_{n,2}^{t_{n-\hat{2},2}}\mathrm{d}\bar{\xi}_{n,1}^{t_{n-\hat{2},1}}\mathrm{d}\zeta_{n,2}^{y_{n-\hat{3},2}}\mathrm{d}\bar{\zeta}_{n,1}^{y_{n-\hat{3},1}}.
    \end{aligned}
\end{align}
By repeating this procedure for all other sites, we define the full tensor $\mathcal{T}_{x_{n}t_{n}y_{n}x_{n-\hat{1}}t_{n-\hat{2}}y_{n-\hat{3}}}$ as
\begin{align}
  \label{eq:GrassmannTensor}
  \mathcal{T}_{x_{n}t_{n}y_{n}x_{n-\hat{1}}t_{n-\hat{2}}y_{n-\hat{3}}}=
  \begin{aligned}[t]
    &T_{x_{n}t_{n}y_{n}x_{n-\hat{1}}t_{n-\hat{2}}y_{n-\hat{3}}}\mathrm{d}\bar{\eta}_{n,2}^{x_{n,2}}\mathrm{d}\eta_{n,1}^{x_{n,1}}\mathrm{d}\bar{\xi}_{n,2}^{t_{n,2}}\mathrm{d}\xi_{n,1}^{t_{n,1}}\mathrm{d}\bar{\zeta}_{n,2}^{y_{n,2}}\mathrm{d}\zeta_{n,1}^{y_{n,1}}\mathrm{d}\eta_{n,2}^{x_{n-\hat{1},2}}\mathrm{d}\bar{\eta}_{n,1}^{x_{n-\hat{1},1}}\\
    &\cdot\mathrm{d}\xi_{n,2}^{t_{n-\hat{2},2}}\mathrm{d}\bar{\xi}_{n,1}^{t_{n-\hat{2},1}}\mathrm{d}\zeta_{n,2}^{y_{n-\hat{3},2}}\mathrm{d}\bar{\zeta}_{n,1}^{y_{n-\hat{3},1}}\left(\bar{\eta}_{n+\hat{1},1}\eta_{n,1}\right)^{x_{n,1}}\left(\bar{\eta}_{n,2}\eta_{n+\hat{1},2}\right)^{x_{n,2}}\\
    &\cdot\left(\bar{\xi}_{n+\hat{2},1}\xi_{n,1}\right)^{t_{n,1}}\left(\bar{\xi}_{n,2}\xi_{n+\hat{2},2}\right)^{t_{n,2}}\left(\bar{\zeta}_{n+\hat{3},1}\zeta_{n,1}\right)^{y_{n,1}}\left(\bar{\zeta}_{n,2}\zeta_{n+\hat{3},2}\right)^{y_{n,2}}.
  \end{aligned}
\end{align}
\end{document}